%% file: Muon_g_2_Experiment_Gray.tex
\documentclass[12pt]{article}
\usepackage{graphicx}
\usepackage{subfigure}

\newcommand\pubnumber{CIPANP2015-Gray \\ FERMILAB-CONF-15-402-E}
\newcommand\pubdate{\today}

\def\regis{Department of Physics and Astronomy\\
Regis University, Denver, CO 80221, USA}
\def\support{\footnote{On behalf of the Fermilab E989 Muon $g$-2 Collaboration}}

\def\Title#1{\begin{center} {\Large #1 } \end{center}}
\def\Author#1{\begin{center}{ \sc #1} \end{center}}
\def\Address#1{\begin{center}{ \it #1} \end{center}}

\newcommand\pubblock{\rightline{\begin{tabular}{l} \pubnumber\\
         \pubdate  \end{tabular}}}
\newenvironment{Abstract}{\begin{quotation}  }{\end{quotation}}
\newenvironment{Presented}{\begin{quotation} \begin{center} 
             PRESENTED AT\end{center}\bigskip 
      \begin{center}\begin{large}}{\end{large}\end{center} \end{quotation}}
\def\Acknowledgements{\bigskip  \bigskip \begin{center} \begin{large}
             \bf ACKNOWLEDGMENTS \end{large}\end{center}}

\input econfmacros.tex

\begin{document}
\begin{titlepage}
\pubblock

\vfill
\Title{Muon $g$-2 Experiment at Fermilab}
\vfill
\Author{ Frederick Gray\support}
\Address{\regis}
\vfill
\begin{Abstract}

A new experiment at Fermilab will measure the anomalous magnetic moment of 
the muon with a precision of 140 parts per billion~(ppb).  This measurement is motivated
by the results of the Brookhaven E821 experiment that were first released more than a decade ago, which reached a precision of 540 ppb.
As the corresponding Standard Model predictions have been refined, the 
experimental and theoretical values have persistently differed by 
about 3 standard deviations.  If the Brookhaven result is confirmed at Fermilab 
with this improved precision, it will constitute definitive evidence for 
physics beyond the Standard Model. The experiment observes the muon spin precession frequency in flight in a well-calibrated magnetic field; the improvement in precision 
will require both 20 times as many recorded muon decay events as in E821 
and a reduction by a factor of 3 in the systematic uncertainties. 
This paper describes the current experimental status as well as 
the plans for the upgraded magnet, detector and storage ring systems that 
are being prepared for the start of beam data collection in 2017.
\end{Abstract}
\vfill
\begin{Presented}
Twelfth Conference on the Intersections \\ of Particle and Nuclear Physics \\
\medskip Vail, Colorado, U.S.A.,  May 19, 2015
\end{Presented}
\vfill
\end{titlepage}
\def\thefootnote{\fnsymbol{footnote}}
\setcounter{footnote}{0}

\section{Introduction}

\begin{figure}[bh]
\centering
\includegraphics[width=\textwidth]{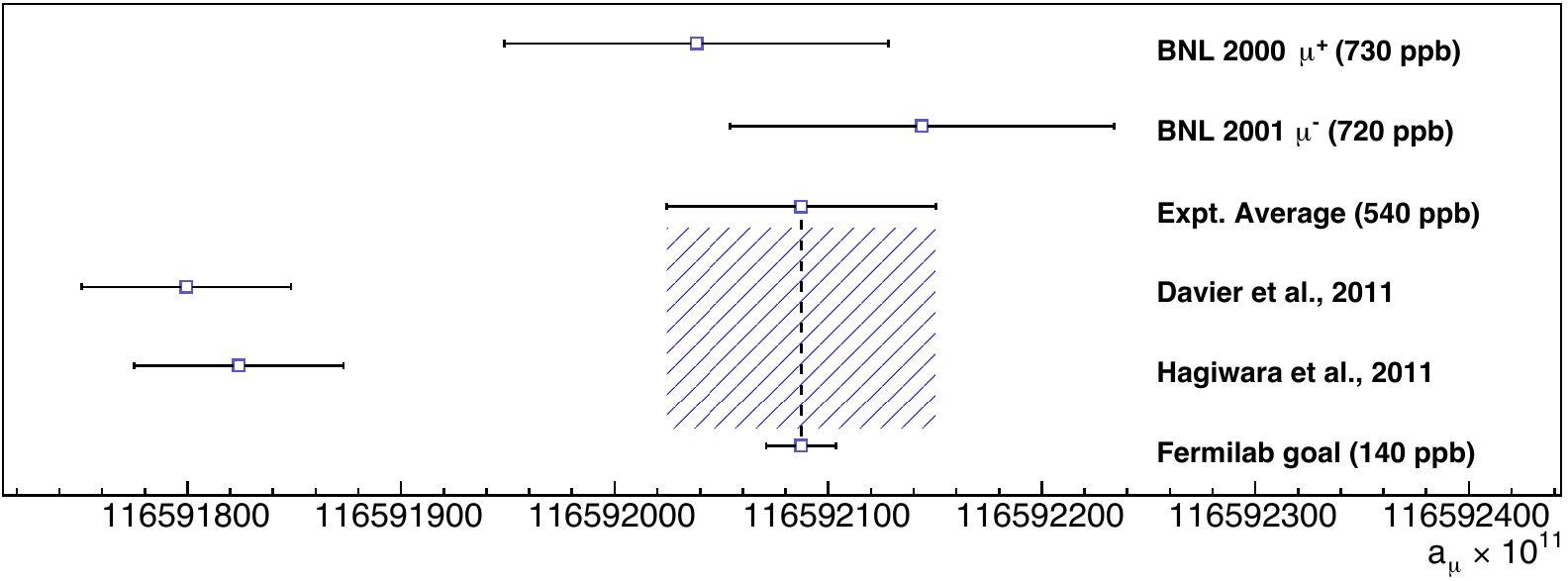}
\caption{Comparison of the Brookhaven E821 results and the projected
sensitivity of the Fermilab E989 Muon $g$-2 experiment to current theoretical
calculations, showing a discrepancy of more than 3 $\sigma$ that could
potentially be due to new physics beyond the Standard Model.}
\label{fig:exptCompare}
\end{figure}

The Fermilab E989 Collaboration is constructing a new experiment to measure
the muon's anomalous magnetic moment, $a_\mu$, with a precision of 140 parts 
per billion~(ppb).  This quantity was last measured by the E821 Collaboration at
Brookhaven National Laboratory~\cite{Bennett:2006fi}.
That experiment reached a final precision of 540~ppb after combining runs with
positive~\cite{Bennett:2002jb} and negative~\cite{Bennett:2004pv} muons,
which were measured to 730~and~720~ppb, respectively.  
As shown in Figure~\ref{fig:exptCompare}, the results disagree by 
more than three standard deviations with recent evaluations of the
theoretical prediction of the Standard Model.  While it is not yet definitive,
the discrepancy strongly suggests that there may be effects on the 
muon's magnetic moment from particles or interactions that are not included
in the Standard Model.

The potential for a discovery of new physics 
provides motivation for the improved measurement of $a_\mu$ at Fermilab.  
Table~\ref{tab:errors} summarizes the planned improvements in statistical
and systematic errors.
Some major systems are being redesigned completely for higher precision,
although a number of major components from E821 are being refurbished and 
reused, particularly the superconducting magnet.
The collaboration includes some veterans of E821, but the number of new
members is much larger, so the experiment 
will effectively provide an independent new measurement.

\begin{table}[bh]
\begin{center}
\begin{tabular}{l|ccc}  
& E821 $\mu^+$ & E821 $\mu^-$   &  Fermilab target  \\ \hline
Statistics  &  620   & 670     & 100      \\ \hline
Systematics in $\omega_a$  & 310      &  210     &  70 \\ 
 {\em{- Overlapping pulses (pile-up)}} & {\em 130} & {\em 80} & {\em 40} \\
 {\em{- Coherent betatron oscillations}} & {\em 210} & {\em 70} & {\em 30} \\
 {\em{- Detector gain changes}} & {\em 130} & {\em 120} & {\em 20} \\
 {\em{- Muon losses}} & {\em 100} & {\em 90} & {\em 20} \\
 {\em{- Others}} & {\em 90} & {\em 100} & {\em 40} \\ \hline
Systematics in $\omega_p$  & 240      &  170     &  70 \\ \hline
Total (quadrature sum) & 730      &  720     &  140 \\ \hline
\end{tabular}
\caption{Projected improvements in statistical and systematic uncertainties, 
in parts per billion.}
\label{tab:errors}
\end{center}
\end{table}
\clearpage

The determination of the theoretical value of $a_\mu = \frac{1}{2} (g-2)$ is 
described in detail in References~\cite{Davier:2010nc, Hagiwara:2011af}.
The Dirac equation predicts that $g=2$, so a nonzero value for $a_\mu$
arises from coupling to virtual particles.
The vast majority (99.6\%) of the value of $a_\mu$ 
arises from the leading-order quantum electrodynamics (QED) process 
that involves the exchange of a single virtual photon,
and higher-order QED contributes nearly all of the rest.  
However, there is essentially no uncertainty 
associated with these contributions; the uncertainty from the hadronic contributions
is dominant.

Approximately three-quarters of the 
squared uncertainty comes from hadronic vacuum polarization, which is 
determined primarily from a 
dispersion integral using hadron production cross section data from $e^+ e^-$
collisions.  
New results from \mbox{BES-III} on the dominant 
$e^+ e^-\rightarrow \pi^+ \pi^-$ process are now available~\cite{Ablikim:2015orh},
and additional measurements with the VEPP-2000 collider in Novosibirsk 
are anticipated~\cite{Rogovsky:2014mka}.  
The remaining error arises from hadronic light-by-light 
scattering, which has historically 
required nuclear modeling techniques.  However, work is in progress to 
calculate it using \mbox{Lattice} QCD~\cite{Jin:2015eua} and with a dispersive 
approach~\cite{Pauk:2014rfa}.   Consequently, the precision of the 
Standard Model value may improve to about 250~ppb based on work that is
already in progress, compared with an uncertainty of 420~ppb today.


The discrepancy of more than three standard deviations could be explained by
a loop diagram describing new particles or interactions.
Although supersymmetry is increasingly constrained by its absence at the 
LHC experiments, it still remains a viable candidate to account for the
discrepancy~\cite{Fargnoli:2013zda,Bach:2015doa}.
As an alternative, additional $U(1)$ gauge bosons known as ``dark photons'' 
were frequently discussed; however, they have 
now been almost excluded as the explanation for the entire
anomaly~\cite{Batley:2015lha}.

The principle of the experiment is to observe the anomalous precession 
frequency $\omega_a$ as a bunched beam of polarized muons circulates in 
an applied magnetic field.  This frequency is the difference between the 
rotation frequencies of the muons' spin and momentum, and it equals 
$a_\mu ( \frac{e}{m} ) B$.
Parity violation causes the high-energy positrons produced in $\mu^+$ decay to 
preferentially follow the 
muon spin direction, so the number of positrons that are 
detected by electromagnetic calorimeters with energy above a defined
threshold is modulated at $\omega_a$.  
This modulation frequency is measured with high precision, as is the 
free proton NMR frequency $\omega_p$ that indicates the magnetic field.
The anomalous magnetic moment is then calculated
as $R/(\lambda - R)$ from the ratio of frequencies $R = \omega_a/\omega_p$ and 
the corresponding ratio of magnetic moments $\lambda = \mu_\mu/\mu_p$.

The storage ring magnet used in E821 was relocated from Brookhaven to Fermilab 
in June and July 2013, and it has now been successfully powered
in its new location.
The circular magnet has a C-shaped cross section and 
consists of a steel yoke and precisely-ground pole pieces that are excited 
by four superconducting niobium-titanium coils; it operates at a field 
of 1.45 T.  At a ``magic'' muon momentum of 3.09 GeV/$c$, the spin precession 
frequency 
is unaffected by electric fields, allowing a large (several kV/cm) electric 
quadrupole field to be used to confine the beam vertically.  The equilibrium 
radius of the beam at this momentum is 7.112~m, and the dilated lifetime
is 64.4~$\mu$s.  Each orbit around the ring requires 149~ns, and the 
anomalous precession period is 4.37~$\mu$s.

\section{Beam and Injection}

The precision of the E821 measurement was limited primarily by statistics.  
The target of the new experiment is a statistical precision 
of 100 ppb, which will require $1.6 \times 10^{11}$ accepted muon decay 
events.  Consequently, a plan has been developed to provide a high-intensity 
muon beam to the experiment.

The Booster at Fermilab delivers 8 GeV protons to the Recycler,
where they will be restructured into 16 bunches per 1.33~s supercycle.
Each bunch will have a full width of $\pm$62~ns.  This primary beam 
will be extracted onto the AP0 target with a separation of 10~ms between 
bunches.  Pions will be collected from the target using a pulsed lithium lens,
and the pion beam will be transported to the Delivery Ring,
which previously served as the antiproton debuncher in the Tevatron era.
Three orbits around this ring will provide an extended path for pions to decay 
in flight into muons, as well as an opportunity to separate the pions and 
muons from proton contamination by time-of-flight.  In all, the pion decay 
path is more than 2~km long, so the result will be essentially a pure muon beam.

The beam will enter the storage ring through a superconducting inflector 
that cancels the main magnet's field.  The inflector, refurbished from E821, 
is a long, narrow pipe with a channel that is $9 \times 28$~mm$^2 \times 1.7$~m.
Design studies are in progress for a possible new inflector with a larger
aperture.  
The beam will then be kicked onto a stored orbit by fast kicker magnets driven 
by a Blumlein pulse forming network.  Unlike
the E821 kicker, which required two turns to 
fully kick the beam, the new system will deliver the entire deflection in the 
first turn.  This kick will populate the phase space of the ring more fully
and will allow the beam to be collimated more completely, reducing muon losses 
at later times in the fill.

The injection process will be optimized not only to reach the maximum stored
intensity, but also to populate the ring's phase space as uniformly and 
symmetrically as possible.  Coherent betatron oscillations arise from the
rotation of the phase space over time at each azimuthal location. 
They are observed as modulations of the centroid position and the
width of the beam.  In E821, the frequency of these oscillations was 
nearly 2$\omega_a$, which directly biased the fit for $\omega_a$.  To move 
significantly above this dangerous frequency, the electric quadrupole system 
is being upgraded to support voltages of 32~kV rather than 24~kV.

Validated models of the beamline and injection process have been developed 
using BMAD and GEANT4.  They predict $7 \times 10^5$ muons per fill at the 
end of the beamline, with a full momentum width $\Delta p/p$ of $\pm 2.5\%$.
This is substantially larger than the ring acceptance of $\pm 0.15\%$, so 
about $2 \times 10^4$ muons per fill will be stored, leading to $1 \times 10^3$
accepted muon decay events with detected positron energies exceeding the 
threshold.  Consequently, it will be possible to collect the required 
statistics within about a year of running time.

\section{Detectors and electronics}

\begin{figure}[tb]
\centering
\includegraphics[width=0.75\textwidth]{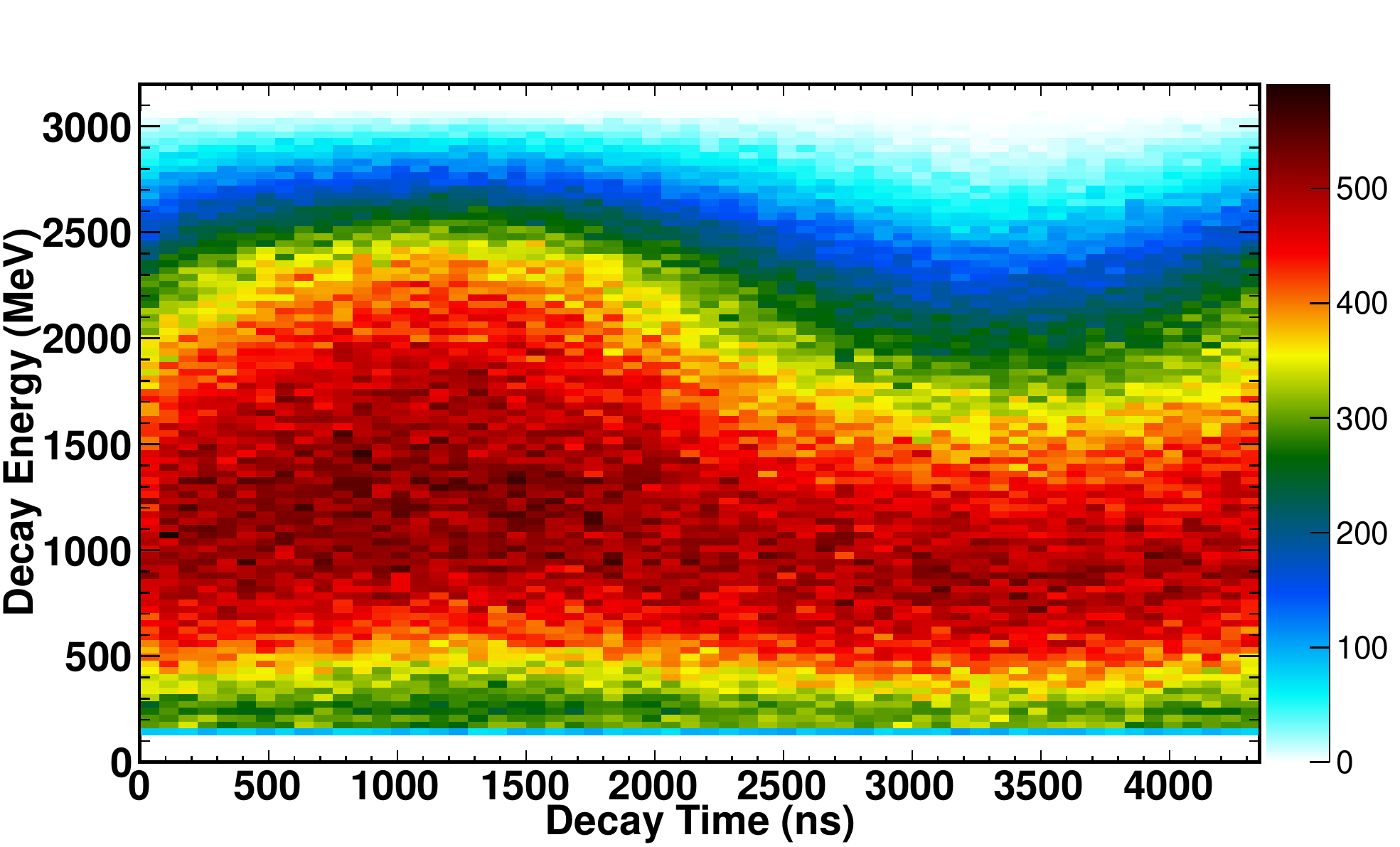}
\caption{Positron energy spectrum as a function of time, shown for one
anomalous precession period.  (Reproduced from \cite{Grange:2015fou}.)}
\label{fig:energy}
\end{figure}

As the muon decays by $\mu^+ \rightarrow e^+ + \nu_e + \bar\nu_\mu$, 
parity violation in the weak interaction causes 
high-energy positrons to follow the muon spin direction.
This correlation is most easily seen by examining limiting cases
in the CM frame; consider that the highest-energy positrons recoil 
back-to-back against the neutrino and antineutrino.  Their fixed helicities
mean that they have zero net spin, so the positron must 
maintain the muon's initial spin.  The effect is then accentuated in the 
transformation to the laboratory frame, where the forward-going
positrons receive an additional forward boost.  
The modulation of the expected positron energy spectrum as the muon spin 
rotates is shown in Figure~\ref{fig:energy}.

In the traditional analysis, a threshold is set at 1.86~GeV,
and the number of positrons exceeding this threshold is recorded.
This choice of threshold maximizes the statistical figure of merit $\sqrt{N} A$,
where $A$ is the modulation asymmetry.
Other analysis methods are possible, and in some cases they have different 
systematic uncertainties.  In one such method, the total deposited energy 
is recorded as a function of time; energy-binned and asymmetry-weighted 
analysis methods can also be explored.

The positrons will be detected by 24 electromagnetic calorimeter
stations spaced equally around the ring.  These stations have been 
designed to minimize the systematic error from ``pile-up'' of 
overlapping pulses from multiple positron hits.  Because higher-energy 
positrons follow a larger-radius arc to the calorimeter, they have a longer
time of flight from the decay point, and therefore they carry a different 
anomalous spin precession phase from lower-energy positrons.  The probability
of mistaking two lower-energy positrons for a single higher-energy one is 
rate-dependent, so it is more likely to happen early in the fill than later.
This time-dependent phase is equivalent to a frequency shift.

Consequently, the calorimeters will be optimized for separation of 
multiple-positron events in both space and time~\cite{Fienberg:2014kka}.
For spatial resolution, they will be segmented into a 9$\times$6 array of 
lead fluoride (PbF$_2$) crystals.
Logistically, this segmentation requires that the crystals be instrumented 
with very compact photodetectors, so Hamamatsu silicon multi-pixel photon 
counters (MPPC) will be used instead of traditional photomultiplier 
tubes (PMT).  The MPPCs are not affected by large magnetic fields, so they do 
not require the long lightguides that would be needed to locate PMTs away from 
the magnet.  Light in PbF$_2$ is produced by the Cerenkov effect, so the width 
of the pulse in time is much faster than in a scintillation-based detector; 
a width of 3~ns (FWHM) has been demonstrated.

To take advantage of the fast pulses, new waveform digitizers are being
constructed.  They will provide 12-bit sampling of each crystal independently
at a rate of 800 MSPS, and they will 
transfer their data to an online computing facility using 
10~GB/s Ethernet.  Every voltage sample from each fill will be available 
to the computer, which will use graphics processing units (GPU) as 
vector processors to extract the times and energies to be recorded.  
This data acquisition system~\cite{Gohn:2015bla} will allow 
the flexibility to implement several analysis methods.

Any time dependence during the fill in the energy response of the detectors 
will also lead directly to a systematic error.  A gain monitoring 
system~\cite{Anastasi:2015ssy} has been developed; it will distribute laser
pulses via a network of splitters and optical fibers to each calorimeter 
segment and to stable monitoring detectors.  This technique has been 
demonstrated to allow measurements of the detector gain to better than 0.1\% 
over the time scale of one fill of the storage ring.

New straw-tube tracking chambers will be placed at two locations in the ring.  
Each will consist of eight U-V straw planes that will allow positrons
to be tracked back to the decay vertex.  The primary role for the trackers
is to measure the muon distribution to use as a weighting in the 
determination of the average magnetic field.  
It will also be used to search for a muon electric dipole moment, which would 
appear as a modulation  of the vertical component of the decay positron 
momentum.  It will also allow the calorimeter to be tested with particles of 
known momentum and with identified pile-up events, and it will allow the 
distribution of lost muons to be characterized.

There will also be a set of four scintillating fiber beam 
monitors that can be plunged into the storage region to measure the
beam profile in both $x$ and $y$ at two locations in the ring, at 180$^\circ$ and
270$^\circ$ from the injection point.
Scintillating fiber monitors are also being designed for installation 
at the entrance and exit of the inflector.  All of these devices can be 
used to optimize and monitor the beam injection.

\section{Magnetic field}

The magnetic field is measured using NMR probes, which are calibrated in terms 
of the free proton precession 
frequency $\omega_p$.  
The field will be continuously monitored by more than 300 fixed probes placed
outside the vacuum chambers. These fixed probes will be calibrated every few 
days to a set of 17 probes housed on an in-vacuum trolley.  The trolley will
in turn be regularly calibrated against an absolute standard probe; a 
spherical H$_2$O-filled probe exists, and an absolute calibration based on
$^3$He is being developed.

The target for the uncertainty in $\omega_p$ is 70~ppb, a factor of 3 
improvement from the level reached in E821.  To reach this target will require 
careful shimming of the magnet, to better than 100~ppb in each multipole.
The stability of the magnet is expected to be excellent in the new MC-1
experimental hall, which has a single-segment concrete floor and which
will provide temperature regulation at the $\pm$2$^\circ$~F level.
Trolley runs will be taken more frequently than in E821, and its
azimuthal position resolution will be improved by a factor of 2. 
The free-induction decay waveforms from all of the fixed probes will be 
recorded, and the temperature dependence of the probes will be calibrated
directly.

\section{Progress and schedule after CIPANP 2015}

At the time of the conference, the reassembly of the storage ring magnet 
at Fermilab was nearly complete.  Only a few final details had to be 
resolved before the magnet could be cooled to 4 K and powered.  
The final beamline magnets were moved into place 
in preparation for field shimming, and civil construction of the external 
beamline enclosures was under way.

In late June, the magnet was successfully ramped up to a current of 3400~A.  
However, an unexpectedly high resistance was measured in one of the lead cans,
requiring that it be re-opened for repair.  Following this repair, the storage 
ring current reached 5300~A on September 22, demonstrating a safety margin 
above the normal operating current of 5100~A that produces a 1.45~T field.  
Consequently, the magnet is now prepared for shimming to begin.  A first scan,
conducted immediately after the first power-up to full current and without 
any adjustments, showed that azimuthal variations in the field were only one 
order of magnitude larger than in the final field map from E821.  

Shimming activity is expected to last until March 2016; at the same time, 
other systems are being prepared for installation: the vacuum chambers, 
the inflector, the kicker, the focusing quadrupoles, and all of the 
field-measuring equipment.  The vacuum
chambers are expected to be installed in April and May 2016, with the remaining
installations of ring-integrated systems following in the summer and fall.
This schedule will provide a period of testing while construction is
completed on the accelerator and beamline systems.  
The first commissioning
of the experiment with beam is expected in March 2017, with production 
data to begin in September after a planned summer shutdown.
It is certainly possible that a first result will be available 
in time for CIPANP 2018.

\Acknowledgements
The author would like to thank the conference organizers for all of their
work; they made CIPANP 2015 an informative and enjoyable event with 
old and new colleagues.  The experiment is supported in part 
by the U.S. Department of Energy and the National Science Foundation.  
Fermilab is operated by Fermi Research Alliance, LLC under contract 
number DE-AC02-07CH11359 with the U.S. Department of Energy.
The author's participation 
at CIPANP 2015 was supported by the National Science Foundation under grant 
number PHY-1206039.

\end{document}

%% file: econfmacros.tex
\def\beq{\begin{equation}}
\def\eeq#1{\label{#1}\end{equation}}
\def\eeqn{\end{equation}}

\def\beqa{\begin{eqnarray}}
\def\eeqa#1{\label{#1}\end{eqnarray}}
\def\eeqan{\end{eqnarray}}

\let\bar=\overbar

\def\Dslash{\not{\hbox{\kern-4pt $D$}}}
\def\dslash{\not{\hbox{\kern-2pt $\del$}}}

\def\msb{{\bar{\ssstyle M \kern -1pt S}}}

